\documentclass[aps,prl,reprint,superscriptaddress,tightenlines,amsmath,amssymb]{revtex4-1} 

\usepackage[utf8]{inputenc} 
\usepackage{graphicx} 
\usepackage{natbib}
\usepackage{mathptmx}
\usepackage{color}
\usepackage[english]{babel}
\usepackage{upgreek}
\usepackage{csquotes}
\usepackage{siunitx}
\DeclareSIUnit\pixel{px}
\usepackage[colorlinks=true,urlcolor=blue,linkcolor=blue,citecolor=blue]{hyperref}

\begin{document}

\title{Anomalous near-equilibrium capillary rise}

\author{Menghua Zhao}
\affiliation{Mati\`{e}re et Syst\`{e}mes Complexes, CNRS UMR 7057, Universit\'{e} Paris Cit\'{e}, 10 Rue A. Domon et L. Duquet, 75013 Paris, France}

\author{Aktaruzzaman Al Hossain}
\affiliation{Department of Mechanical Engineering, Stony Brook University, Stony Brook, NY 11794, USA}

\author{Carlos E. Colosqui}
\email[E-mail address: ]{carlos.colosqui@stonybrook.edu}
\affiliation{Department of Mechanical Engineering, Stony Brook University, Stony Brook, NY 11794, USA}
\affiliation{Department of Applied Mathematics and Statistics, Stony Brook University, Stony Brook, NY 11794, USA}
\affiliation{Institute for Electrochemically Stored Energy,  Stony Brook University, Stony Brook, NY 11794, USA}

\author{Matthieu Roch\'{e}}
\email[E-mail address: ]{matthieu.roche@u-paris.fr}
\affiliation{Mati\`{e}re et Syst\`{e}mes Complexes, CNRS UMR 7057, Universit\'{e} Paris Cit\'{e}, 10 Rue A. Domon et L. Duquet, 75013 Paris, France}


\begin{abstract}
We report and rationalize the observation of a crossover from the classical Lucas-Washburn dynamics to a long-lived anomalously slow regime for capillary rise in simple glass tubes. We propose an analytical model considering the role of thermal motion and the nanoscale surface topography to account for the experimental observations. The proposed model indicates that the contact line perimeter and the surface topography dimensions determine the crossover condition and anomalous imbibition rate. Our findings have important implications for the scientific understanding and technical application of capillary imbibition and suggest strategies to control the adsorption of specific liquids in porous materials.
\end{abstract}
\maketitle

The phenomenon of capillary rise is among the most studied in interfacial science owing to its relevance to numerous natural and industrial processes such as liquid transport in porous media \cite{morrow1970physics,lu2004rate,dang2005characterization,gruener2009capillary}, the wetting of fabrics \cite{ferrero2003wettability,duprat2022moisture}, additive manufacturing \cite{arrabito2019imbibition,azhari2017binder}, and microfluidic devices for liquid handling \cite{li2019bioinspired,anderson2022automated}. 
The classical theoretical description of capillary rise relies on mechanical models involving inertia, viscous hydrodynamic friction, gravitational effects and the driving capillary force. According to Jurin's law \cite{jurin1718,rodriguez2010derivation}, these forces produce mechanical equilibrium when the rising liquid reaches the equilibrium height $h_{eq}=2\gamma \cos\theta_Y/(\rho g R)$
with $\gamma$ the surface tension of the liquid, $\theta_Y$ the Young contact angle, $\rho$ the density of the liquid, $g$ the acceleration of gravity, and $R$ the radius of the tube. The surface of the solid is conventionally assumed as smooth and chemically homogeneous, while fluid and solid phases are separated by sharp interfaces.
 
The general analytical treatment of capillary rise is complex when considering short-time inertial effects \cite{quere1997inertial,fries2008analytic,fries2008transition,das2012early,das2013different}. However, as the column height $h=h(t)$ increases and inertial effects become negligible, the system enters a regime described by the celebrated Lucas-Washburm (LW) equation \cite{lucas1918ueber, washburn1921dynamics}
\begin{equation}
    \dot{h}=\frac{\rho g R^2}{8\mu} (h_{eq}/h-1),
    \label{Eq:dh} 
\end{equation}
where $\mu$ is the liquid viscosity.
A key assumption of this description is that the contact angle is constant and equal to the equilibrium contact angle $\theta_{eq}=\theta_Y$, as predicted by Young's law. While the validity of this assumption is not trivial, the LW equation accounts accurately for experimental observations for small imbibition rates $\dot{h}\ll\gamma/\mu$ \cite{mumley1986,popescu2008capillary,li2013experimental,heshmati2014experimental,walls2016capillary,wu2017capillary}.

Anomalous behavior with large deviations from LW predictions via Eq.~\ref{Eq:dh}
has been reported in porous media with random networks of micro-scale pores \cite{delker1996interface,lago2001capillary}. This anomalous behavior, characterized by vanishing imbibition rates, has been attributed to the random non-local dynamics of the wetting front, contact line pinning, and spatial fluctuations of the capillary pressure, which are ignored in the LW equation \cite{delker1996interface,lago2001capillary,ganesan1998dynamics,geromichalos2002nonlocal,alava2004imbibition}. 
Notably, various interfacial phenomena such as colloidal particle adsorption, shear-driven drainage, and droplet spreading \cite{kaz2012physical,colosqui2013colloidal,colosqui2016crossover,rahmani2016colloidal,keal2018colloidal,jose2018physical}  also show an anomalously slow relaxation to equilibrium after a regime crossover from the initial non-equilibrium dynamics. 
These observations are attributed to surface energy barriers induced by surface topography at the nanoscale.

To the best of our knowledge, no prior study has characterized and rationalized the observation of anomalous capillary rise in a capillary tube. A  small number of studies report that prewetting is needed to observe agreement with LW predictions \cite{mumley1986,walls2016capillary}.  
Here, we report that capillary rise experiments performed over several hours in capillary tubes display a predictable crossover from conventional LW dynamics to an anomalously slow rise near equilibrium. To rationalize this phenomenon, we propose an analytical description based on a Langevin-type equation accounting for thermal motion and energy perturbations induced by the nanoscale surface topography. 
The model produces close agreement with our experimental observations and the magnitude of its parameters is supported by topographic analysis of the surface of the capillaries via atomic force microscopy (AFM).

\begin{figure}[!htb]
    \centering
    \includegraphics[width=1.\columnwidth]{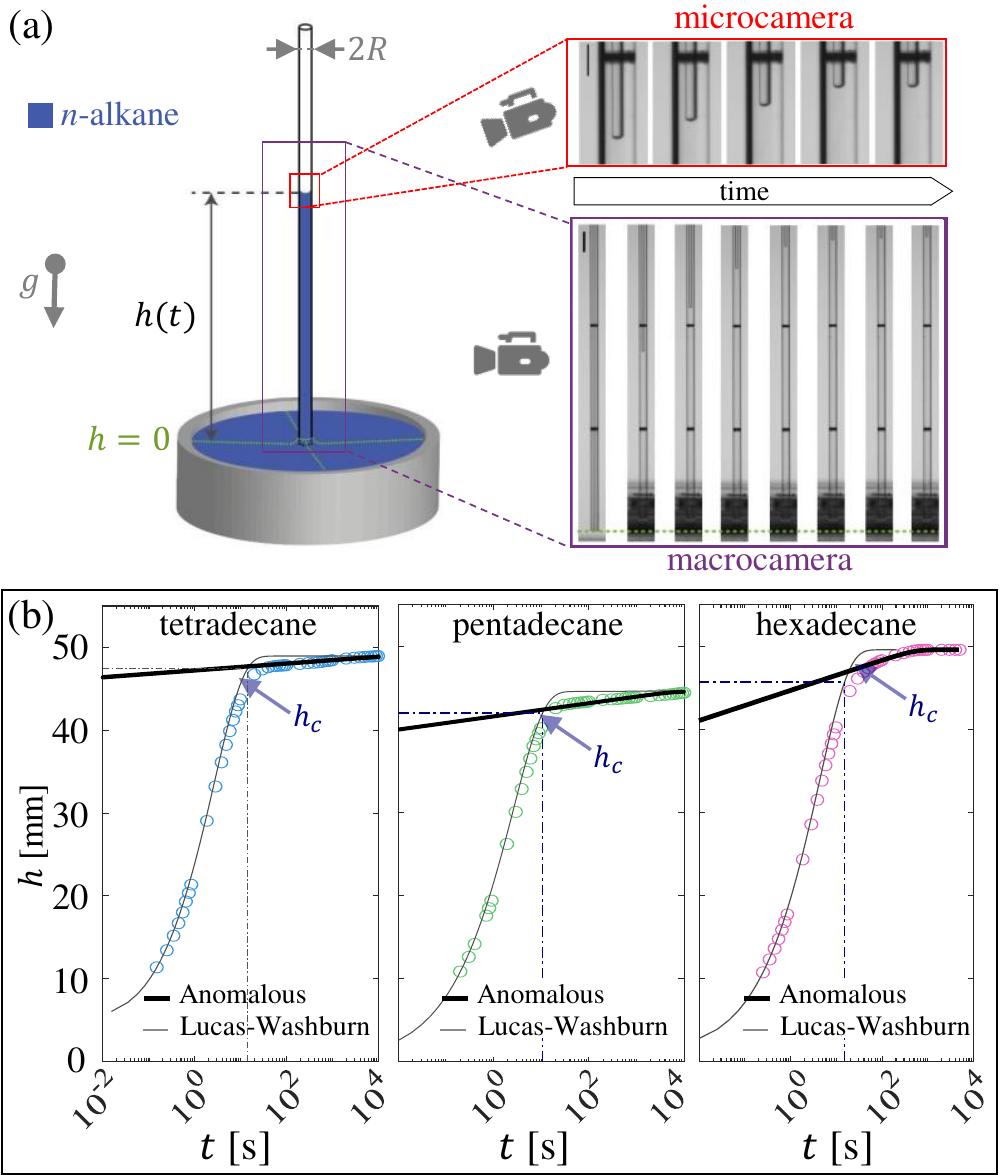}
    \caption{Capillary rise in a glass microcapillary. 
    (a) Experimental setup. Image sequences are for pentadecane in the 140-\si{\micro\meter} microcapillary. Top panel: microcamera view at $t=$ 26.5, 266.5, 2646.5, 26496.5 s and 20 hours; scale bar: 1 mm. Bottom panel: macrocamera view at $t=$ -10, 2, 4, 8, 16, 32, 64 and 128 s; scale bar: 3 mm. Dashed green line: position of the flat bath surface. (b) Column height $h(t)$ and regime crossovers for three alkanes in the 140-\si{\micro\meter} microcapillary. Analytical predictions (see legend) for the crossover height $h_c$ (Eq.\,\ref{Eq:hc}), LW dynamics (Eq.\,\ref{Eq:dh}), and anomalous rise regime (Eq.\,\ref{Eq:hK}) employ the parameters reported in Table \ref{Tab:1}.}
    \vskip -1em
    \label{fig:fig1}
\end{figure}

We analyze the case of a liquid with constant mass density $\rho$, surface tension $\gamma$, viscosity $\mu$, and temperature $T$ that rises within a vertical capillary of radius $R$ much smaller than the capillary length $\ell_c=\sqrt{\gamma/(\rho g)}$ (Fig.\,\ref{fig:fig1}a). 
The liquids in this study are three homologous alkanes with low vapor pressures (Sigma Aldrich, purity 99\%; tetradecane 
($\text{C}_{14}\text{H}_{30}\geq$ 99\%, $\rho=762$ kg/m$^3$, $\gamma=26.1$ mN/m, $\mu=2.1$ mPa s, and $P_{vap}=2.0$ Pa at 25\si{\celsius}),
pentadecane 
($\text{C}_{15}\text{H}_{32}\geq$ 99\%, $\rho=769$ kg/m$^3$, $\gamma=27.1$ mN/m, $\mu=2.3$ mPa s, and $P_{vap}=0.7$ Pa at 25$^{\circ}$C), 
and hexadecane 
($\text{C}_{16}\text{H}_{34}\geq$ 99\%, $\rho=770$ kg/m$^3$, $\gamma=27.5$ mN/m, $\mu=3.4$ mPa s, and $P_{vap}=0.2$ Pa at 25$^{\circ}$C)). 
We assume thermal and chemical equilibrium with the much less viscous ambient air. 
We use glass capillaries (Hirschmann ringcaps 5 $\mu $L and 20 $\mu $L) with two different inner radii, $R=$ \SI{140}{\micro\meter} and $R=$ \SI{310}{\micro\meter} (5\% uncertainty). Capillaries are used as received, without precleaning or prewetting. Liquids are poured into a borosilicate glass petri dish (Pyrex, diameter 30 mm). A lid with a tight hole to insert the capillary tube covers the dish to prevent contamination and further minimize evaporation of the nonvolatile liquid during long exposure to the ambient. 
All experiments are performed at a controlled temperature $T=25\pm1$ \si{\celsius}.
Further details on the experimental protocol and reproducibility  are included in the Supplemental Material \cite{SM}.

A \enquote{microcamera} (Imaging Source, DFK camera) records the displacement of the meniscus at the top of the liquid column with a spatial resolution of 6.0 \si{\micro\meter\per\pixel} (top right panel in Fig.\,\ref{fig:fig1}a). We use a sub-pixel technique \cite{laurence2012tracking} to detect displacements as small as 1.2 \si{\micro\meter}.
A \enquote{macrocamera} (Andor Zyla 5.5) captures an overall view of liquid rise from the initial contact to the final equilibrium height with a resolution of 42.2 \si{\micro\meter\per\pixel} (bottom right panel in Fig.\,\ref{fig:fig1}a). Images are processed with the software package FiJi \cite{schindelin2012} to extract the column height $h(t)$ measured from the flat bath interface to the bottom of the rising meniscus as a function of the time $t$ elapsed since contact.
Figure \ref{fig:fig1}b shows the column height evolution $h(t)$. At early times, the three alkanes show a fast increase of the column height that slows down as the system approaches equilibrium, which we defined at the height $h_{eq}$ for which there is no detectable change for at least 6000 s.
While experiments agree closely with the LW model (Eq.\,\ref{Eq:dh}) for typical observation times $t\lesssim 10$~s, substantial discrepancies arise for longer times. In particular, the time to reach the experimentally determined equilibrium height is over two orders of magnitude larger than that predicted by the LW equation.

To rationalize our experimental observations we will consider the effects of thermal fluctuations and energy perturbations induced by nanoscale surface topography. The time evolution of the liquid height $h(t)$ is governed by a Langevin-type equation \cite{colosqui2013colloidal,colosqui2016crossover,COLOSQUI2018654}
\begin{equation}
\xi \dot{h}=-\frac{1}{\xi}\frac{\partial {\cal F}}{\partial h}+\sqrt{2k_B T\xi}\eta(t),
\label{Eq:L} 
\end{equation}
where $\xi=8\pi\mu h$ is the damping coefficient assuming energy dissipation is dominated by Poiseuille flow in the narrow capillary, $\eta(t)$ is a zero-mean unit-variance uncorrelated noise, and the system free energy is
\begin{equation}
{\cal F}=\frac{\rho{g}\pi{R^2h^2}}{2}
-2\pi{R}\gamma\cos\theta_{Y} h
+\frac{\Delta U}{2}\sin\left(\frac{2\pi h}{\ell}+\phi\right).
\label{Eq:F} 
\end{equation}
The first and second terms in Eq.~\ref{Eq:F} are, respectively, the gravitational potential energy and the interfacial free energy determined by the Young contact angle $\theta_Y$, which both appear in the LW equation (Eq.~\ref{Eq:dh}). The third term is a single-mode energy perturbation of magnitude $\Delta U$, period $\ell$, and phase $\phi$, induced by nanoscale surface topography and/or chemical heterogeneities. 
Since $\ell\ll h_{eq}$, we can arbitrarily adopt $\phi=-2\pi h_{eq}/\ell$ so that mechanical equilibrium is exactly attained at the height $h_{eq}$ for which the system energy in Eq.\,\ref{Eq:F} is at the global minimum, in accordance with Jurin's law.

Any finite energy perturbation of magnitude $\Delta U>0$ in Eq.~\ref{Eq:F} produces multiple metastable states where $\partial{\cal F}/\partial h=0$, sufficiently close to equilibrium, when $\lvert h_{eq}-h\rvert \times\rho g R^2 \ell/2\le \Delta U$.
Hence, as $h\to h_{eq}$ the system governed by Eqs.~\ref{Eq:L}-\ref{Eq:F} must eventually crossover from non-equilibrium dynamics driven by capillary forces to arrested dynamics dominated by metastable state transitions induced by thermal fluctuations. 
Such a regime crossover corresponds to the transition from conventional LW dynamics to an anomalous capillary rise and occurs around the crossover height \cite{colosqui2013colloidal}
\begin{equation}
h_c=h_{eq}-\alpha\frac{\Delta{U}}{\rho g R^2 \ell},
\label{Eq:hc} 
\end{equation}
where a factor $\alpha=$~0.5 estimates the center of the range of heights over which the crossover takes place \cite{colosqui2013colloidal,rahmani2016colloidal,colosqui2016crossover,keal2018colloidal,jose2018physical}.

For $h \gtrsim h_c$, the column height evolution is determined by the displacement rate $\dot{h}=\ell (\Gamma_+-\Gamma_-)$. Using Kramers theory for the (forward/backward) transition rates $\Gamma_\pm$ gives \cite{colosqui2013colloidal}
\begin{equation}
\dot{h}=\frac{h_{eq}}{h}V_H\sinh\left(\frac{h_{eq}-h}{L_H}\right)
\label{Eq:dhK} 
\end{equation}
with the ``hopping'' length
\begin{equation}
L_H=\frac{2k_B T}{K\ell},
\label{Eq:LH} 
\end{equation}
and the ``hopping'' velocity 
\begin{equation}
V_H=
\frac{\ell}{2\pi\xi_{eq}}
\sqrt{\frac{1}{4}\left(\frac{2\pi}{\ell}\right)^4\Delta{U}^2-K^2}
\exp\left(\frac{-\Delta{U}-\frac{1}{8}K\ell^2}{k_B T}\right),
\label{Eq:VK} 
\end{equation}
where $K=\rho g \pi R^2$ and $\xi_{eq}=8\pi\mu h_{eq}$ is the damping coefficient prescribed by the equilibrium height.
The displacement rate in Eq.~\ref{Eq:dhK} can be integrated to obtain the implicit relation $t=-(L_H/V_H)\times [F(x)+G(x)]+c$, where 
$x=(h-h_{eq})/L_H$, 
$c$ is an integration constant,
$F(x)=(1-x L_H/h_{eq})\times\log[\tanh(x/2)]$, and 
$G(x)=(L_H/h_{eq})\times[\mathrm{Li}_2(e^{-x})-\mathrm{Li}_2(-e^{-x})]$; here, $\mathrm{Li}_2$ is the dilogarithm function.
For $L_H\ll h_{eq}$, which is expected for the case of nanoscale surface features, the expression for the column height can be simplified to
\begin{equation}
t=-\frac{L_H}{V_H} \log\left[\tanh\left(\frac{h_{eq}-h}{2L_H} \right)\right].
\label{Eq:hK} 
\end{equation}

The near-equilibrium expressions for the displacement rate (Eq.~\ref{Eq:dhK}) and column height (Eq.~\ref{Eq:hK}) derived via Kramers theory are prescribed by the energy barrier magnitude $\Delta U$ and period $\ell$.
Assuming that free energy perturbations in the 1D energy profile ${\cal F}(h)$ (Eq.~\ref{Eq:F}) are caused by nanoscale topographic features with a characteristic projected area $A_d$, we then define the period $\ell=A_d/(2\pi R)$ and energy barrier $\Delta U=\beta \gamma A_d$, where $\beta$ is a shape factor accounting for surface energy changes associated with different 3D interfacial configurations induced by the nanoscale topography   \cite{colosqui2013colloidal,colosqui2016crossover,COLOSQUI2018654}.
We use $A_d$ and $\beta$ as fitting parameters to account for our experimental observations of (i) the critical crossover height $h_c$ (Eq.\,\ref{Eq:hc}) and (ii) the displacement rate $\dot{h}$ (Eqs.\,\ref{Eq:dhK}-\ref{Eq:hK}) for $h>h_c$. 
We estimate the Young contact angle $\theta_Y\simeq\theta_{eq}$ from the measured equilibrium heights. 
Table\,\ref{Tab:1} reports the values of the parameters employed in the model.

\begin{figure*}[!htb]
	\centering
	\includegraphics[width=.95\textwidth]{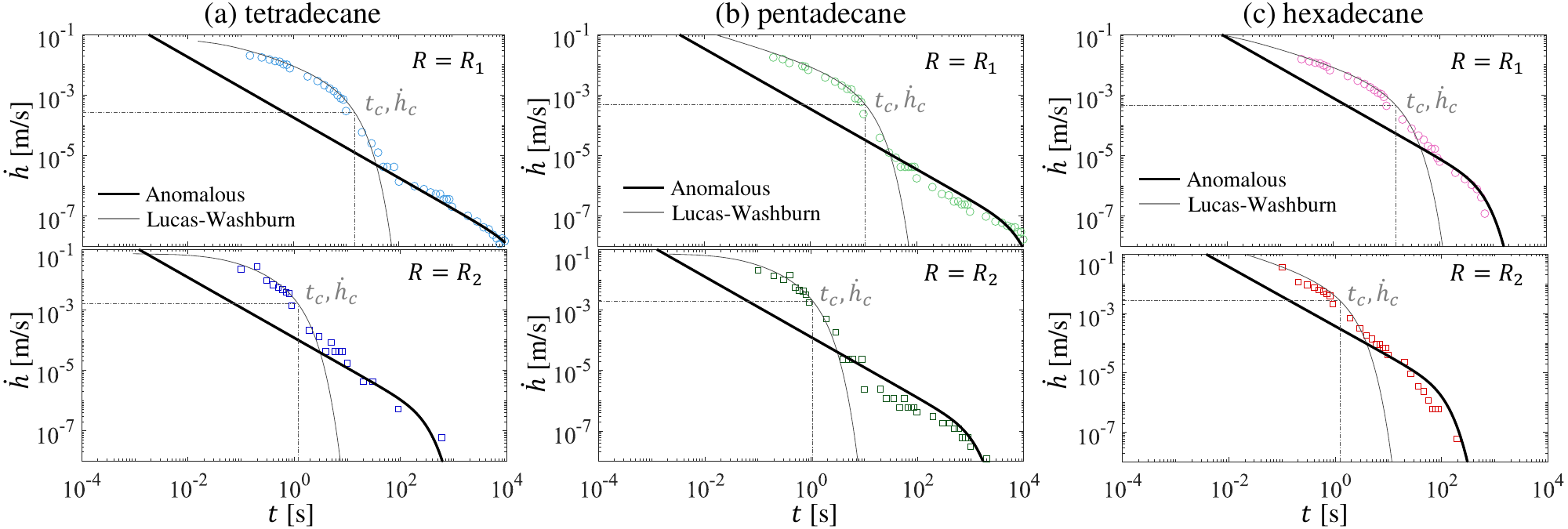}
\vskip -8 pt	
\caption{Capillary rise rates $\dot{h}\equiv dh/dt$ over observation times $10^{-1}\leq t\leq 10^4$ s in two glass microcapillaries of radius $R_1=140$ \si{\micro\meter} and $R_2=310$ \si{\micro\meter} for (a) tetradecane, (b) pentadecane, and (c) hexadecane. Symbols: experimental data. Thin continuous line: Eq.~\ref{Eq:dh} (LW dynamics); thick continuous line: Eq.~\ref{Eq:dhK} (anomalous regime). The crossover rate $\dot{h}_c$ and time $t_c$ (dot-dashed lines) are obtained by employing the crossover height $h_c$ (Eq.~\ref{Eq:hc}) in Eqs.~\ref{Eq:dh}.}
    \vskip -1em
\label{fig:figure2}
\end{figure*}

The positive displacement rates $\dot{h}$ reported in Fig.~\ref{fig:figure2} decay monotonically over time with a marked transition from LW dynamics to the anomalous capillary rise regime near equilibrium.
For reference, the maximum evaporation flux into a perfect vacuum, prescribed by the vapor pressure and molecular weight of the studied alkanes, translates into negative rising rates $\dot{h}\simeq$~-6 to -1 \si{\micro\meter\per\second}, a value that is higher by many orders of magnitude than the actual evaporation rate \cite{beverley1999evaporation}.
While the LW model (Eq.\,\ref{Eq:dh}) describes the initial capillary rise dynamics, the proposed model (Eqs.\,\ref{Eq:dhK}-\ref{Eq:hK}) accounts for the late near-equilibrium evolution of $\dot{h}$ by using the model parameters reported in Table \ref{Tab:1}. 
The same finding is reported in Fig.\,\ref{fig:fig1}b for the column height evolution.
The reported values of $A_d$ and $\beta$ (cf. Table \ref{Tab:1}) give the characteristic energy barriers $\Delta U \simeq$ 6.7 to 10.7 $k_BT$ and periods $\ell\sim 10^{-14}$~m. 
We note that $\ell$ in our model is determined by the average displacement of the full contact line perimeter over a single surface feature of nanoscale area $A_d$ and therefore it is much smaller than the physical distance between localized surface features \cite{colosqui2013colloidal}.

\begin{table}[h]
\begin{center}
\begin{tabular*}{0.915\columnwidth}{lcc|c|cc|c}
\hline
Case $R_1$ & $h_{eq}$ [mm] & $\theta_{eq}$ [$^\circ$] & $A_d$ [nm$^2$] 
& $\Delta U/k_B T$ & $\ell$ [pm] & $\theta_c$ [$^\circ$]\\
\hline
$\text{C}_{14}\text{H}_{30}$ &  49.0 & 11.3 & 86.75 & 10.67 & 0.099 & 18.2\\
$\text{C}_{15}\text{H}_{32}$ &  44.6 & 15.5 & 45.04 & 9.71 & 0.051 & 24.9\\
$\text{C}_{16}\text{H}_{34}$ &  49.6 & 17.7 & 21.25 & 6.73 & 0.024 & 28.6\\
\hline
\end{tabular*}
\vskip 3pt
\begin{tabular*}{0.915\columnwidth}{lcc|c|cc|c}
\hline
Case $R_2$ & $h_{eq}$ [mm] & $\theta_{eq}$ [$^\circ$] & $A_d$ [nm$^2$] 
& $\Delta U/k_B T$ & $\ell$ [pm] & $\theta_c$ [$^\circ$]\\
\hline
$\text{C}_{14}\text{H}_{30}$ &  22.0 & 12.3 & 59.64 & 8.69 & 0.031 & 19.8\\
$\text{C}_{15}\text{H}_{32}$ &  20.9 & 14.0 & 55.72 & 10.11 & 0.029 & 22.5\\
$\text{C}_{16}\text{H}_{34}$ &  22.2 & 19.2 & 18.68 & 6.96 & 0.009 & 31.1\\
\hline
\end{tabular*}
\end{center}
\vskip -15 pt
\caption{Parameters employed in the analytical predictions for the two microcapillaries of radius $R_1=140$ \si{\micro\meter} and $R_2=310$ \si{\micro\meter}. The shape factor $\beta=1-\cos\theta_Y$ is determined using the estimate $\theta_Y=\theta_{eq}$.}
\vskip -1em
\label{Tab:1}
\end{table}

An important feature of the proposed analytical model is the prediction of the conditions for the crossover from LW dynamics to the anomalous regime.
The crossover heights $h_c$ predicted via Eq.\,\ref{Eq:hc} are in good agreement with the heights at which the transition is observed experimentally (i.e., 5\% to 10\% below $h_{eq}$). 
The use of Eq.\,\ref{Eq:hc} in Eq.\,\ref{Eq:dh} gives the corresponding critical displacement rate $\dot{h}_c$ and time $t_c$ reported in Fig.~\ref{fig:figure2}. 
Besides, the proposed analytical model indicates that the contact angle $\theta_c=\arccos(\rho g R h_c/(2\gamma))$ at the crossover height is significantly larger than $\theta_{eq}$ (cf. Table \ref{Tab:1}),
which highlights the importance of following the rising dynamics over sufficiently long times to determine equilibrium properties.

\begin{figure}[!htb]
	\centering
	\includegraphics[width=0.98\columnwidth]{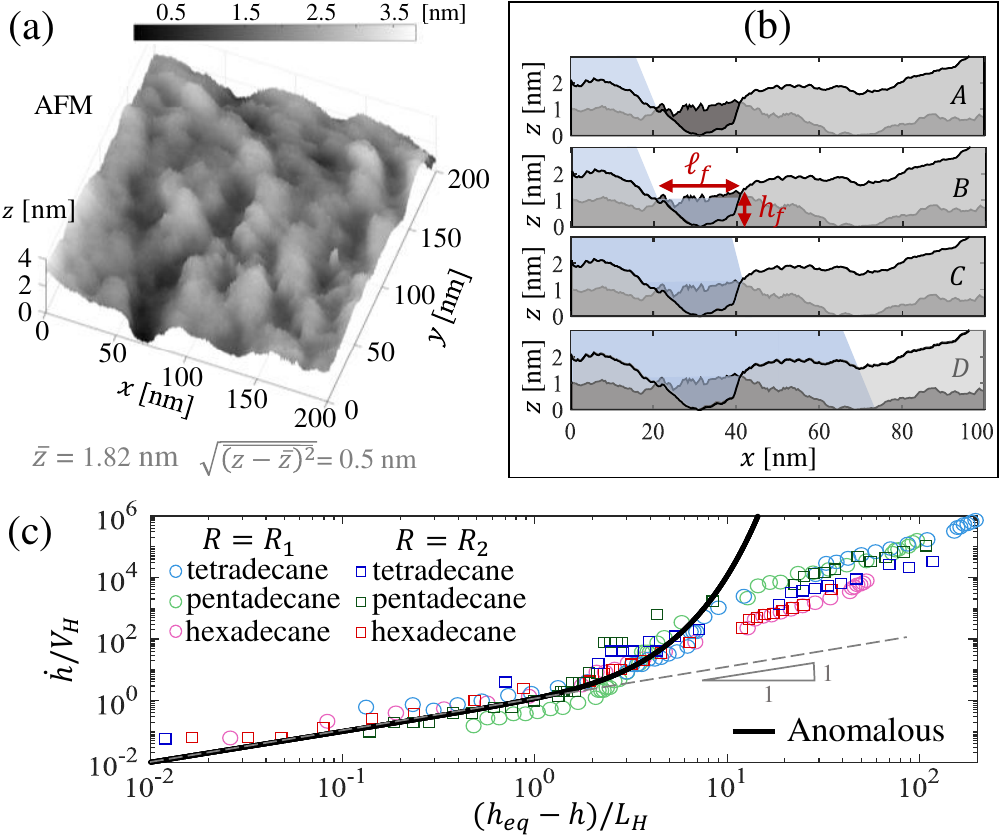}
\vskip -8 pt	
	\caption{Nanoscale surface topography and master curve in the stochastic regime.
(a) AFM image of the surface topography height $z(x,y)$.
(b) Illustration of contact line motion over surface features with characteristic extent $\ell_f$ and height $h_f$ using two overlapping AFM scans along the x-direction 10 nm apart in the y-direction.
The sequence labeled (A) to (D) shows the hypothesized infiltration of critically small features via hemiwicking.   
(c) Dimensionless displacement rate $\dot{h}/V_H$ versus separation from equilibrium $(h_{eq}-h)/L_H$ for all tested systems. Thick black line: non-dimensionalized theoretical prediction. Grey dashed line: linear scaling in the limit $(h_{eq}-h)<L_H$.}
\vskip -1.5em
	\label{fig:figure3}
\end{figure}

The close agreement between experiments and the proposed model relies on a set of parameters  (cf. Table \ref{Tab:1}) that we assume to be induced by topographic surface features with a specific range of nanoscale dimensions. To verify this assumption we image the nanoscale topography of the inner surface of the capillaries at several different locations with an atomic force microscope (Park NX-20, $k=$~42 N/m, $\omega_o\!=\!$~330 kHz). The topographic height data $z(x,y)$ (Fig.\,\ref{fig:figure3}a), obtained in non-contact mode with a lateral resolution of 0.39 nm and height noise level of 0.03 nm, reports a mean topographic height $\bar{z}=1.82$ nm, standard deviation $z_{std}=\overline{\sqrt{z-\bar{z}}}=0.501$ nm, and a nearly Gaussian height distribution (kurtosis $\kappa =$ 2.96).
The AFM images (cf. Fig.\,\ref{fig:figure3}a) show that the wetting front must move over a a complex landscape of surface features with a wide range of lateral  dimensions $\ell_f \sim$~10 to 100 nm and a characteristic height $h_f\simeq 2 z_{std}\simeq 1$ nm (Fig.\,\ref{fig:figure3}a).   

We now turn to the analytical estimation of the energy barriers $\Delta U$ induced by the topographic features imaged via AFM. Based on the energetics of wetting on a ``rough'' surface \cite{bico2002wetting}, hemiwicking (i.e., liquid infiltration in the local topography) is favored by features with lateral dimensions $\ell_f\le\ell^*$ prescribed by the critical infiltration length  
$\ell^* = h_f\times (1/\cos^2\theta_Y-1)^{-1/2}$.
As illustrated in Fig.\,\ref{fig:figure3}b, a local displacement of the contact line over a liquid-infiltrated feature with characteristic projected area $A_d=\pi {\ell^*}^2/4$ produces a characteristic energy barrier $\Delta U=\Delta U^*=\gamma (1-\cos\theta_Y) A_d$, from which we readily determine the shape factor $\beta=1-\cos\theta_Y$. 
Notably, by using $h_f=$~0.85 to 1.05 nm and the experimental estimates for the equilibrium contact angle $\theta_{eq}$ we predict $\ell^*\simeq$ 5 to 10 nm and the values for $A_d$ and $\beta$ reported in Table I that are employed in the analytical fits reported in Fig.~\ref{fig:fig1}b \& Fig.~\ref{fig:figure2} for the anomalous regime.  
It is worth noting that for surface features with lateral dimensions $\ell_f > \ell^*$ larger than the critical infiltration length, the ``dry'' topography (Fig.\,\ref{fig:figure3}b) induces energy barriers given by a shape factor $\beta\simeq \sqrt{1+(h_f/\ell_f)^2}-1\sim (h_f/\ell_f)^2$ and thus we have $\Delta U\ll \Delta U^*$. Hence, the near-equilibrium displacement rate (Eq.\,\ref{Eq:dhK}) is determined by the nanoscale features that are infiltrated by the preceding liquid film.

Based on this analysis, the characteristic ``hopping'' rate $V_H$ and length $L_H$ are prescribed by the critical surface feature length $\ell_f\simeq \ell^*$. The functional form of Eq.~\ref{Eq:dhK} suggests a collapse of the experimental results onto a single curve when plotting the dimensionless displacement rate $\dot{h}/V_H$ against the equilibrium separation $(h_{eq}-h)/L_H$ (see Fig.~\ref{fig:figure3}c).
Furthermore, sufficiently close to equilibrium, $(h_{eq}-h)<L_H$, we find the linear scaling $\dot{h}=V_H\times(h_{eq}-h)/L_H$ (cf. Fig.~\ref{fig:figure3}c) with the characteristic displacement rate in the range $V_H\simeq$~0.06 to 4.3 $\mu$m/s for the anomalous regime.
It is worth remarking that these findings and the model in Eq.~\ref{Eq:F} are strictly valid for sufficiently large separation of scales between the characteristic dimensions of the surface topography and the radius of the capillary tube so that $\ell\ll h_{eq}$ and $h_f\ll R$. While this condition is satisfied for microcapillaries with macroscopically smooth and flat surfaces having nanoscale topographic features, the proposed model is not applicable for surface features with dimensions comparable to the capillary radius.

In summary, capillary rise experiments performed over unusually long times with non-volatile liquids under controlled ambient conditions report a crossover from LW dynamics to an anomalous regime as the column height approaches equilibrium.
Compared to the classical LW equation, the characteristic defect area $A_d$ is the only additional parameter in the model proposed to account for the experimental observations. The estimate $\theta_Y=\theta_{eq}$ for the Young contact angle results in a shape factor $\beta=1-\rho g R h_{eq}/(2\gamma)$ with an error linearly proportional to the uncertainty in determining $h_{eq}$. While the equilibrium height was determined with high accuracy owing to the extended observation times, even large relative uncertainties up to 20\% in $h_{eq}$ would result in slightly different defect areas $A_d$ that produce the energy barrier $\Delta U$ and period $\ell$ reported in Table 1. We thus find that the proposed model could similarly fit the experimental data with a reasonable single estimate for the Young contact angle and small adjustments of the reported nanoscale defect areas $A_d$.

Our experimental and theoretical analyses thus suggest that the crossover height and the anomalous imbibition rate are determined by topographic surface features with nanometric dimensions to favor hemiwicking and infiltration of liquid preceding the contact line. Our analysis indicates that tuning the base radius-to-height ratio of nanoscale topographic features can promote or prevent the occurrence of the anomalously slow imbibition. In addition, we find that the equilibrium contact angle can be determined from the linear relation between the displacement rate and separation from equilibrium in the final stage of the anomalous capillary rise. These findings have direct implications on the design and use of capillary devices for micro/nanofluidic handling, the characterization of porous materials, and the fundamental understanding of capillary driven transport in near-equilibrium conditions. 

\begin{acknowledgements}
C.\,C.\,acknowledges Universit\'{e} Paris Cit\'{e} for supporting his stay at Mati\`{e}re et Syst\`{e}mes Complexes and support from the NSF (CBET-2016204). M.\,Z.\,and M.\,R.\, gratefully acknowledge ANR (Agence Nationale de la Recherche) and CGI (Commissariat à l’Investissement d’Avenir) for their financial support of this work through Labex SEAM (Science and Engineering for Advanced Materials and Devices), ANR-10-LABX-0096 and ANR-18-IDEX-0001.
\end{acknowledgements}

\section{Supplemental Information}
\subsection{Experimental reproducibility}

\begin{figure}[]
	\centering
	\includegraphics[width=0.8\columnwidth]{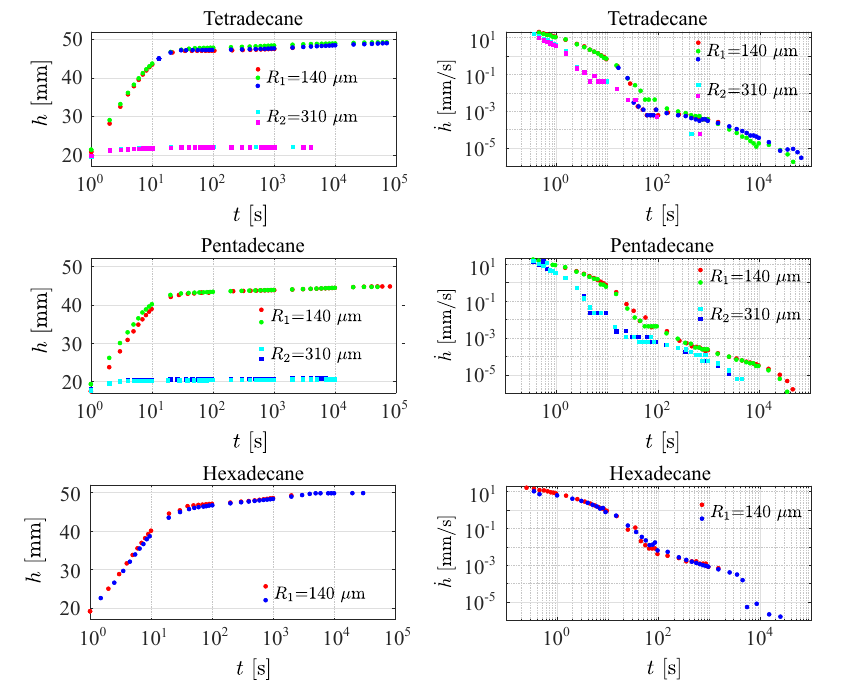}
	\caption{Reproducibility of the capillary rise experiments. The glass capillary in these experiments are employed as received, without prewetting with the alkanes or pre-cleaning. Different color markers (see legend) corresponds to different realizations of the same experimental conditions in a new capillary of rdius $R_1=140$~$\mu$m or $R_1=310$~$\mu$m.
}
	\label{fig:figureS1}
\end{figure}

Following the experimental protocol described in the main text, we performed multiple realization of the capillary rise experiments for the three studied alkanes to verify the reproducibility of the observed anomalous regime. 
For these experiments the capillary tubes (Hirschmann ringcaps 5 $\mu $L and 20 $\mu $L) were employed as received from the supplier without prewetting with the alkanes or pre-cleaning.
The column height $h(t)$ and displacement rate $\dot{h}$ measured in these experiments are reported in Fig.~\ref{fig:figureS1} for the two different capillary sizes employed (iner radius $R_1=$~140~$\mu$m and $R_2=$~310~$\mu$m) and show a high degree of reproducibility.
For the experiments reported in Fig.~\ref{fig:figureS1}, the room temperature was controlled at $T=25.5\pm 1$ \textcelsius\,for the tetradecane and pentadecane, and $T=22.0\pm 1$\textcelsius~ was the temperature for the experiments with hexadecane.

\subsection{Cleaning protocol and surface aging}

\begin{figure}[!hb]
	\centering
	\includegraphics[width=0.8\columnwidth]{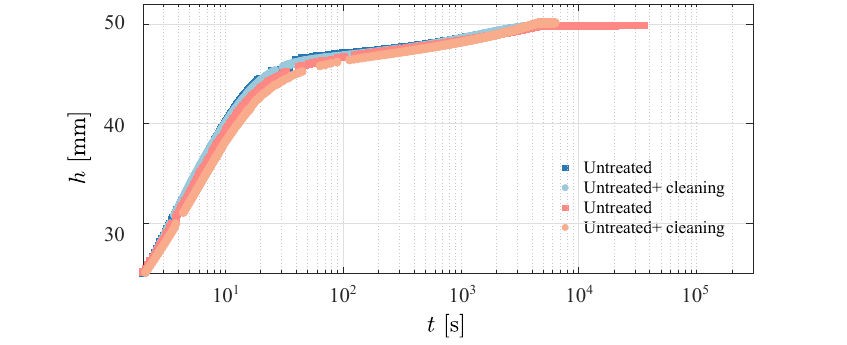}
	\caption{Effects of pre-cleaning and surface aging. 
Different marker colors (see legend) correspond to different capillary rise experiments with hexadecane using a new capillary untreated or pre-cleaned.
The ``untreated'' cases correspond to capillary tubes that were not exposed to ambient air and were employed without a cleaning protocol, as received from the supplier. 
The ``pre-cleaned'' cases correspond to capillary tubes that were treated with the cleaning protocol before the experiments. 
The ``aged'' cases correspond to capillary tubes that were employed without pre-cleaning after a long exposure to ambient air for 1 and 4 months.
The room temperature is $T=$~22.0$\pm$ 1\textcelsius~ for all the reported cases.
}
\label{fig:figureS2}
\end{figure}

To assess the effect of pre-cleaning the glass capillary, a set of capillary rise experiments were performed by employing a cleaning protocol.
The cleaning protocol consisted of injecting successively ethanol, acetone, and distilled water through the capillaries using a syringe pump at a fixed flow rate of 100 $\mu$L/min. Each of the three cleaning steps was performed for 10 minutes. The capillary was finally dried with pressurized nitrogen. 
The column height $h(t)$ observed with and without the cleaning protocol is reported in Fig.~\ref{fig:figureS2} and shows no substantial differences on the observed rise dynamics and final equilibrium height.  

The effect of the aging of the hydrophilic glass surface was additionally examined by exposing the glass capillaries to the ambient air over periods of 1 to 4 months before performing the experiments without employing a cleaning protocol (cf. Fig.~\ref{fig:figureS2}).
The capillaries are made of hydrophilic (DURAN) borosilicate glass, which makes them susceptible to surface aging, with the formation of oxides and contamination.
As reported in Fig.~\ref{fig:figureS2}, the aging of the surface on capillaries not treated with cleanin protocol has a significant effect on the observed rise dynamics and final equilibrium height. 

%

\end{document}